\documentclass[twocolumn, secnumarabic,amssymb, nobibnotes, aps, prx]{revtex4-2}

\usepackage[utf8]{inputenc}
\usepackage{lipsum}
\usepackage{amsmath}
\usepackage{graphicx}
\usepackage{bigints}
\usepackage{soul}
\usepackage{verbatim}
\usepackage{natbib}

\newcommand{\half}{\frac{1}{2}}
\newcommand{\qtr}{\frac{1}{4}}

\newcommand{\dg}{\dagger}
\newcommand{\pd}{\partial}

\newcommand{\la}{\langle}
\newcommand{\ra}{\rangle}
\newcommand{\ld}{\lambda}
\newcommand{\Ld}{\Lambda}
\newcommand{\E}{{\cal E}}

\newcommand{\pb}[1]{\la \psi_{#1}|}
\newcommand{\pk}[1]{| \psi_{#1} \ra}
\newcommand{\pbi}{\pb{i}}

\newcommand{\pkj}{\pk{j}}
\newcommand{\poii}{\pk{i}\pb{i}}

\newcommand{\pojj}{\pk{j}\pb{j}}
\renewcommand{\v}[1]{\mathbf{#1}}
\newcommand{\vx}{\v{x}}
\newcommand{\vk}{\v{k}}
\newcommand{\vq}{\v{q}}
\newcommand{\vp}{\v{p}}
\newcommand{\vs}{\v{s}}

\newcommand{\ee}{\mathrm{e}}

\renewcommand{\i}{\ensuremath{\operatorname{i}}}
\newcommand{\tr}{\ensuremath{\operatorname{tr}}}
\renewcommand{\Re}{\ensuremath{\operatorname{Re}}}

\renewcommand{\d}[1]{\ensuremath{\operatorname{d}\!{#1}\,}}
\newcommand{\td}[1]{\ensuremath{\frac{\operatorname{d}\!{#1}}{\operatorname{d}\!t}\,}}
\newcommand{\vd}[1]{\ensuremath{\operatorname{d^3}\!{#1}\,}}
\begin{document}
\title{Quantum Signatures of Gravity from Superpositions of Primordial Massive Particles}

\author{Gowtham Amirthya Neppoleon$^{1,2}$}
\email{ganeppoleon@connect.ust.hk}
\author{Aditya Iyer$^3$}
\email{aditya.iyer@physics.ox.ac.uk}
\author{Vlatko Vedral$^{3,4,5}$}
\email{vlatko.vedral@physics.ox.ac.uk}
\author{Yi Wang$^{1,2}$}
\email{phyw@connect.ust.hk}
\affiliation{%
$^1$Department of Physics, The Hong Kong University of Science and Technology, Clear Water Bay, Kowloon, Hong Kong, P.R.China
}%
\affiliation{%
$^2$The HKUST Jockey Club Institute for Advanced Study, Clear Water Bay, Kowloon, Hong Kong, P.R.China
}%
\affiliation{%
$^3$Townsend Laboratory, Department of Physics, University of Oxford, Oxford OX1 3PU, United Kingdom
}%
\affiliation{
$^4$Centre for Quantum Technologies, National University of Singapore, Block S15, 3 Science Drive 2, Singapore
}%
\affiliation{
$^5$Department of Physics, National University of Singapore, 2 Science Drive 3, Singapore 117542}
\date{\today}

\begin{abstract}
    We study the superposition of primordial massive particles and compute the associated decoherence time scale in the radiation dominated universe. We observe that for lighter primordial particles with masses up to \(10^7\,\rm{kg}\), the corresponding decoherence time scale is significantly larger than the age of the observable universe, demonstrating that a primordial particle would persist in a pure quantum state, with its wavefunction spreading freely. For heavier particles, they can still be in a quantum state while their position uncertainties are limited by the wavelength of background photons. We then discuss three observational signatures that may arise from a quantum superposition of primordial particles such as primordial black holes and other heavy dark matter candidates, namely, interference effects due to superpositions of the metric, transition lines in the gravitational wave spectrum due to gravitationally bound states indicating the existence of gravitons, and witnesses of quantum entanglement between massive particles and of the gravitational field.
\end{abstract} 
\maketitle

The reconciliation of quantum theory with general relativity (GR) has been a longstanding open problem in physics \cite{issue1,issue2,issue3,issue4}. A central tenant of quantum mechanics (QM) is that quantum degrees of freedom can be superposed. A straightforward extension of this principle to a massive particle opens up several issues that are at the heart of the conflict between GR and QM. Preserving the unitarity and linearity of quantum mechanics, we expect a massive superposition to manifest as a superposition of the metric itself which if observed would be a tell-tale sign of quantum gravitational effects. 

In this work we study primordial massive particles, that are relics from the early universe. Examples of primordial massive particles include dark matter candidates that may be created in the reheating era or primordial black holes (PBHs) that are sourced by primordial fluctuations that re-enter the horizon. We consider a particular system comprising a primordial particle, interacting with the photon background in the radiation dominated universe and demonstrate that it can exist in a superposition. Furthermore, we compute the decoherence timescale associated with such a state and conclude that the superposition can persist up to present times for a range of masses. 
The gravitational decoherence of light dark-matter for well-chosen initial states through the study of scattering processes with the environment was explored in \cite{DMSuporposeNewtonian}. This has also been extended to the context of general relativistic scattering \cite{DMSuperposePRL, DMSuporposeGR}. While Ref. \cite{DMSuperposePRL} has considered decoherence by a single scattering particle in terms of the decay of the system's off-diagonal density matrix elements, with a special focus of light dark matter with coherent oscillations, we consider heavy dark matter and decoherence due to a thermal environment of particles interacting with the system through gravity. This analysis is performed for an arbitrary initial state of the system. Furthermore, we investigate the implications of a coherent massive superposition to the quantum nature of gravity. In particular, we outline observational consequences relating to interference patterns produced by such superpositions, quantum gravitational bound states, and entanglement witnesses.

\textit{Coherence of the Superposition} --- The massive particle interacts continuously with matter and radiation in the universe. In the following, we will consider decoherence due to photons as a representative example of this phenomenon. We model the massive particle by a scalar field \(\chi\) which interacts with the photon bath, described by the Lagrangian,
\begin{equation} \label{eqn:free-lagrangian}
    \mathcal{L} =  \half \pd_\mu\chi^\dg(\vx) \cdot  \pd^\mu \chi (\vx) - \half M^2 \chi^2 + \qtr F^{\mu\nu}(\vx) F_{\mu\nu}(\vx)-V,
\end{equation}
where \(M\) is the mass of the massive particle. In this equation and in what follows, we will employ natural units, \( \hbar = c = G = 1\). In these units, the gravitational potential of interaction, \( V \) between these fields is,
\begin{equation} \label{eqn:hint-pos}
V = \int \vd \vx \vd \vx' \frac{M}{|\vx - \vx'|}  \chi^\dg(\vx)  \chi(\vx)\, \varrho(\vx').
\end{equation}
Here, \(\varrho(\vx)\) is the photon energy density at \(\vx\) which is defined in terms of the energy of single photon, \( \epsilon_\vk = k \) as,
\begin{equation} \label{eqn:rho-photon}
\varrho(\vx) = \int \vd \vk \; \ee^{\i \vk \cdot \vx}\, \epsilon_\vk\, b_\vk^\dg b_\vk.
\end{equation}
The decoherence of the state of the massive particle is studied in Appendix A in terms of the decay of \( \tr \rho^2 \) where \(\rho\) is its reduced density matrix. We begin by considering the Fourier transformed interaction Hamiltonian,
\begin{align}
    H_\text{int} &= \int  \vd \vp \vd\vk\,  \nu(\vk)\, \epsilon_\vk\, a_\vp^\dg a_{\vp + \vk}\, b_\vk^\dg b_\vk,
\end{align}
where \( \nu(k) = M/\pi k^2 \) is the Fourier transform of gravitational potential \( \phi(\vx) = M/|\vx| \) (see Appendix B). Let \( \rho_T\) be the density matrix of the system and environment combined, and \(\rho_\E\) be the reduced density matrix of the environment. The unitary evolution of \( \rho_T \) under \( H_\text{int}\) interaction is given by, $\rho_T(t) =  \exp{(-\i H t)}\, \rho_T (0)\, \exp{(\i H t)}$. Since we are interested in the description of the primordial particle (in terms of its reduced density matrix \(\rho\)), we trace out the photon environment and obtain the well-known Lindblad form for the master equation \cite{Halliwell2007},
\begin{align}
\td \rho  \Delta t &= \i [ \tr_\E (U_1 \rho_{\E}) - {\tr}_\E ( B \rho_\E), \rho ] + \nonumber \\
&\qquad \tr_{\E} \left(  U_1 \rho_T U_1 - \half U_1^2 \rho_T - \half \rho_T U_1^2 \right),
\end{align}
where \( U_1 = - \int_{-\infty}^{\infty} \d t H_\text{int}(t) \) is the time evolution operator, and \( B \) is some Hermitian operator which drops out in the one particle sector. Here, \( \Delta t \) is the timescale over which we study the evolution of \(\rho\). This timescale is small compared to the evolution of the system but large compared to the evolution of the environment. The right hand side of this equation is also proportional to \(\Delta t\), details of which can be found in Ref. \cite{Halliwell2007}.

We now represent the density matrix as the function \( \rho(\vk, \vk') := \la \vk | \rho | \vk' \ra \), and compute the rate of change of \( \tr \rho^2 \) as an indicator of the purity of the state. We obtain, 
\begin{equation}
    \td{(\tr\rho^2)}(\vk,\vk') =  -\frac{1}{\pi} \int \vd \vq \, |\nu(q)|^2 \, \epsilon_\vq^2\, n_\vq (n_\vq +1)\, \Lambda (q),
\end{equation}
where \(\Lambda(q)\) is defined by,
\begin{equation}
    \Lambda (q) := \tr \rho^2 - \Re  \int \vd\vk \vd\vs  \rho(\vk,\vs) \rho(\vs-q\, \hat{z},\vk-q\, \hat{z}).
\end{equation}
Having shown that \(\Lambda(q) \in [0, \tr\rho^2]\), we obtain,
\begin{equation}
    \td{(\tr\rho^2)}(\vk,\vk') = -\Gamma \tr \rho ^2,
\end{equation}
where the decay rate \( \Gamma \in [0,\Gamma_0] \), with,
\begin{equation} \label{eqn:general-gamma-0}
    \Gamma_0 = \frac{4M^2}{\pi^2} \bigintssss_0^\infty \frac{\d q}{q^2}  \epsilon_\vq^2 n_\vq (n_\vq + 1).
\end{equation}
Further details of this calculation can be found in Appendix A. Setting \( \epsilon_\vq = q \), and \(n_\vq\) to be the Planck number density, the integral can be explicitly evaluated to yield,
\begin{equation} \label{eqn:gamma-0}
   \Gamma_0  = \Bigg[\Bigg(\frac{16}{15 \pi^2} - \frac{96 \zeta(5)}{\pi^6}\Bigg)\frac{1}{\beta^5} + \frac{8 \zeta(3)}{\pi^4} \frac{1}{\beta^3}\Bigg] M^2.
\end{equation}
where \(\zeta(x)\) is the Riemann zeta function.
Thus, we have derived an upper bound on the rate of decoherence of the massive particle due to the photons that scales as the squared mass.

The result in \eqref{eqn:general-gamma-0} is also generally applicable to decoherence caused by the baryonic matter with \( \epsilon_\vq = m^2 \) and \( n_\vq \) taken to be the Fermi-Dirac distribution \( (1+\exp{\beta q^2/2m})^{-1} \) in the non-relativistic limit. However, in this case the integrand in \eqref{eqn:general-gamma-0} is \( \propto q^{-2} \) in the IR limit, and hence the integral diverges. We resolve this issue by using a better upper bound than \( \tr \rho^2 \) in the derivation of \eqref{eqn:general-gamma-0}. We show in Appendix C that indeed in the IR limit, this introduces a factor \( \propto q^2 \) that alleviates the divergence. It must be noted that this argument also holds for photons and will yield a better bound \(\Gamma_0 \) in place of \eqref{eqn:gamma-0}. However, we shall neglect this effect in the photon case for simplicity.

\begin{figure}
    \centering
    \includegraphics[width=\columnwidth]{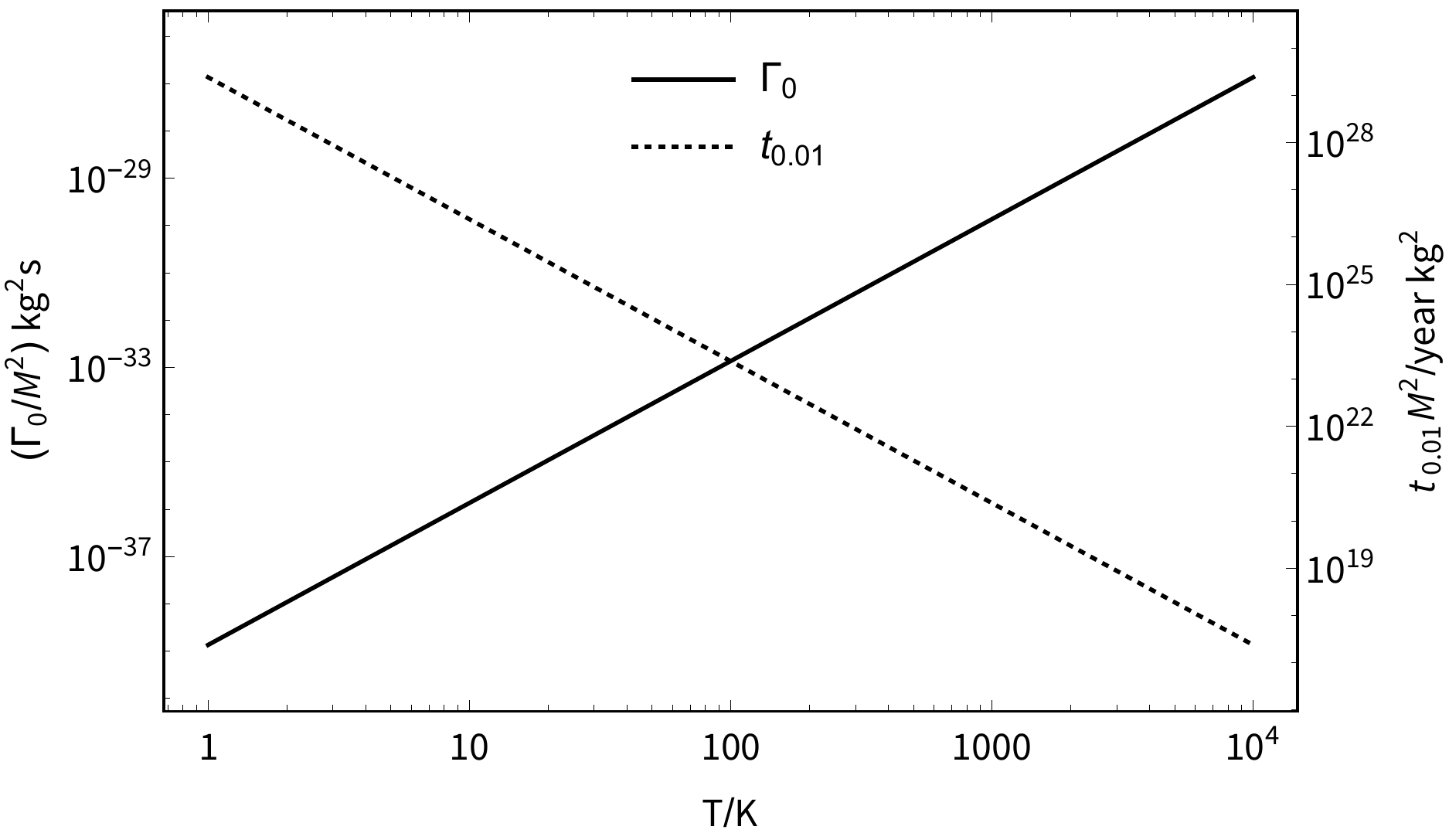}
    \caption{Plots of the maximum decoherence rate \( \Gamma_0 \) and time taken for 1\% decoherence due to background photons \( t_{0.01} \) of a  1 kg primordial particle as a function of the temperature of the universe. Note that \( \Gamma_0 \) scales as \( M^2 \) while \( t_{0.01} \) scales as \( M^{-2} \).}
    \label{fig:gammas}
\end{figure}

We observe in Fig. \ref{fig:gammas} that for a range of temperatures spanning the current temperature (2.7 K) of the universe to that at the time of recombination (3000 K) the decoherence rate is mild enough such that even a single percentage drop in purity (the corresponding time taken is \( t_{0.01} = -\ln (0.99)/\Gamma_0 \)) occurs at timescales several orders greater than the age of the universe for  \(M = 1 \mathrm{kg}\). Considering the \(M^{-2}\) dependence of this time, we can safely neglect decoherence for particles at a mass scale of order \( 10^7\, \rm{kg}\) or lower. While we can extend our discussion to more massive candidates of dark matter, such as Massive Compact Halo Objects (MACHOs) \cite{Macho1, Macho2, Macho3}, they decohere rather quickly as calculated above. However, it is still possible to observe effects of quantum spread as their localization in position can only be of order of the wavelength of the CMB photons. 

Similarly, we also compute the minimum spread associated with a primordial particle in the time when decoherence is negligible. For demonstration, we consider an initial Gaussian state of the primordial particle with initial spread $s_0$. After time $t$ the spread of the wavefunction is,
\begin{equation}
    s(t) = \sqrt{\frac{ s_0^4 + (\hbar t/m)^2 }{ s_0^2}}.
\end{equation}
This can be minimized in \(s_0\) to give the minimum value of \(s(t)\) to be,
\begin{equation} \label{eqn:minimal-spread}
    s_\text{min}(t) = \sqrt{\frac{2\hbar t}{m}}.
\end{equation}
For instance, for a primordial particle of mass \(10^{11}\, \)GeV formed during early universe, \(t \approx 14\) billion years and we get the minimum spread today as \( s_\text{min} \approx 0.72\,\)m. Note that this is only the minimal wave function spread considering the uncertainty principle. The initial condition of the particle may allow more dramatic spread of the wave function, depending on the production mechanism of the primordial particle.

\textit{Observable Consequences} --- Having shown that a class of massive particles do not decohere significantly since their formation through the interaction with photons in the radiation dominated universe, we will discuss how a superposition of massive particles and its QG effects can be observed. We will consider three classes of observations --- (i) Experiments that distinguish between classical spreads and quantum superpositions of the stress energy tensor, (ii) stationary states and transitions, and (iii) signatures of entanglement. 

The quantum nature of gravity, unlike standard field theory, comprises two distinct aspects --- whether the metric can exist in a superposition, and whether there is a quantized force carrier for gravity -- the graviton. Of the three proposed signatures, the first allows us to probe the quantum nature of the metric. The second signature explores the existence of atom-like bound states for superpositions with a discrete absorption spectrum hinting at the existence of gravitons. Finally, detecting entanglement between the components of the bipartite system interacting gravitationally demonstrates that the gravitational field mediating the interaction is quantum.

\textit{Superposition Effects} --- One of the key questions we wish to answer about the nature of quantum gravity is how superpositions of matter affect the construction of the stress-energy tensor and in turn affect the associated gravitational field. In a semi-classical treatment of QFT in curved space time, one might consider a background value of the stress energy tensor which is in actuality an expectation value over possible quantum states of the universe, effectively treating the stress energy tensor classically. Alternatively, in a fully quantum approach, one may consider a many-worlds scenario -- one for each distribution of matter and all of them superposed linearly.

As an illustration consider the following -- It is well known that massive objects distort the path of passing light around them giving rise to gravitational lensing. We investigate how this phenomenon will occur with a massive object (such as primordial black hole (PBH) or a heavy particle) with a spread in wave function. Consider a detector on Earth that measures the angular intensity \( I(\theta, \phi) \) of radiation coming from different azimuths \( \theta \) and declinations \( \phi \). Let \( T_{\mu\nu} \) be the stress energy tensor associated with the superposed massive object with its center described by a wavefunction \( \psi(\vx) \), and let \( \alpha(\theta, \phi, T_{\mu\nu}) \) be the corresponding angular amplitude of radiation from a distant source, lensed by \( T_{\mu\nu} \) and observed on Earth in that direction. Then, we may compute a semi-classical average \(T_{\mu\nu}\) defined by,
\begin{equation}
    \la T_{\mu\nu}(\vx) \ra = \int \vd \vx' |\psi(\vx')|^2 \tau_{\mu\nu}(\vx-\vx'),
\end{equation}
where \( \tau_{\mu\nu}(\vx) \) is the the energy momentum tensor at \( \vx \) of the massive object centered at the origin. In the semi-classical picture, we will expect to observe a lensing pattern due to this expectation value, \(\la T_{\mu\nu} \ra\) whose intensity \(I_{cl} \) is given by,
\begin{equation}
\label{eq:cl-superpose}
    I_\text{cl}(\theta, \phi) = \Big| \alpha\Big(\theta, \phi, \la T_{\mu\nu} \ra\Big) \Big|^2.
\end{equation}
However, we can alternatively consider a coherent sum over the amplitudes given by,
\begin{equation} \label{eq:qg-superpose}
    I_\text{qg}(\theta, \phi) = \Big| \int \vd \vx' \psi(\vx') \alpha(\theta, \phi, S_{\vx'}\tau_{\mu\nu} )  \Big|^2,
\end{equation}
where \( S_{\vx'}\tau_{\mu\nu} \) is the shifted energy momentum tensor, \( S_{\vx'}\tau_{\mu\nu}(\vx) = \tau_{\mu\nu} (\vx + \vx') \). We will expect these intensities \( I_\text{cl} \) and \( I_\text{qg} \) to be qualitatively different which would in turn be an observable signature. \( I_\text{cl} \) would be equivalent to the lensing pattern produced by an extended object, while \( I_\text{qg} \)  would show the emergence of a diffraction pattern corresponding to the interference of Einstein rings generated by each individual branch of the PBH superposition. 

This effect resembles Feynman's thought experiment \cite{CH1} to witness the quantum effects of gravitational field through a double slit experiment with single particles in presence of gravitational field. It has been argued that this does not prove the existence of non-commuting complementary observables of the gravitational field \cite{BMV1} and thus does not prove that the gravitational field is quantum in the quantum information theoretic sense. However, in this work, we have demonstrated that a cosmological massive particle can persist in a superposition. Thus, in the case of a coherent sum over amplitudes in \eqref{eq:qg-superpose}, we are adding geodesics corresponding to different stress energy tensors, rather than only considering the effect of gravitational field on the phase. While, this is still not conclusive evidence of quantum nature in the quantum information theoretic sense, we expect the interference pattern to be more exotic than a simple double slit pattern that could be reasonably produced by an alternative classical mechanism. A more rigorous signature of the quantum nature of the gravitational field could be constructed by considering the role of entanglement as discussed subsequently under entanglement effects.

\textit{Stationary States and Transitions} --- A quantum system of heavy massive particles, possibly belonging to the dark matter sector can now exist in a quantum bound state \cite{Gravatom1,Gravatom2}. As an illustration, consider a pair of such particles of mass \( M \) bound by mutual gravitation. We expect that this system would not emit gravitational waves as long as it is in a stationary quantum state. However, similar to atoms, this system would absorb and emit gravitational waves during transitions. 

First, we compute the spectrum of energies by quantizing the total angular momentum similar to the Bohr atom. The total energy of a state with principal quantum number \( n \) is,
\begin{equation} \label{eqn: energy-spectrum}
    E_n = - \frac{G^2 M^5}{4\hbar^2 n^2} = - 1.84 \times 10^{-32} \, \Bigg( \frac{M c^2}{\mathrm{10^{11}\,GeV}} \Bigg)^5\, \frac{1}{n^2}\  \mathrm{J}.
\end{equation}
The corresponding orbit radius from the common center of mass is,
\begin{equation}
    r_n = \frac{\hbar^2n^2}{GM^3} = 29.1 \ \Bigg(\frac{\mathrm{10^{11}\,GeV}}{M c^2}\Bigg)^3\,n^2\, \mathrm{pm}.
\end{equation}
Note that the radius of this orbit is much smaller than the minimal uncertainty in position of a \(\mathrm{10^{11}\,GeV}\) mass, calculated using \eqref{eqn:minimal-spread}. Similarly, the velocity of the particle in orbit is given by,
\begin{equation}
    v_n = \frac{Gm^2}{2n\hbar} = 10.1 \, \Bigg( \frac{M c^2}{10^{11}\,\mathrm{GeV}} \Bigg)^2\, \frac{1}{n}\  \mathrm{(nm/s)}.
\end{equation}
Note that this approximation is only valid when \( v_n \ll c \), or correspondingly \( M \ll 10^{19}\, \mathrm{GeV/c^2} \), which is expected generally of dark matter particles. Imagine the scenario when a gravitational wave passes through a cloud of these bound states. As the spectrum of energies is discrete, we would expect to see absorption lines similar to the Hydrogen atom spectrum. The frequencies of such lines can be expressed in terms of the principal quantum numbers of initial and final states \( n \) and \( m \) as,
\begin{align}
    \nu_{nm} &= \frac{E_0}{\hbar} \Bigg(\frac{1}{m^2} - \frac{1}{n^2} \Bigg) \nonumber \\
    &= 174 \, \Bigg( \frac{M c^2}{10^{11}\,\mathrm{GeV}} \Bigg)^5\, \Bigg(\frac{1}{m^2} - \frac{1}{n^2} \Bigg)\, \mathrm{Hz}.
\end{align}

The gravitational cross section for absorption has been calculated to be of the order \( l_p^2 \) \cite{Gravatom1}. Although this is very small, we can consider the augmenting effect of dark matter prevalent over astronomical distances. While we may not observe very dark lines like the EM atoms, we could expect dips in the GW spectrum. 
Furthermore, this effect will depend on whether we can get a small enough line width, corresponding to a large intensity drop at specific frequencies. Despite the significant observational challenge this poses, it is worthwhile to investigate this possibility thoroughly as it is among the few known direct indicators of the graviton.

\textit{Entanglement Effects} ---
Having shown that a primordial massive particle can exist in a coherent quantum state, it is natural to expect the emergence of entanglement in superpositions of composite systems. For instance, similar to the entanglement between the proton and electron of the hydrogen atom, constituent primordial particles of binary systems will also be entangled. 

The entangled pair would undergo interference which is markedly different from the unentangled case \cite{BMV1,BMV2}. However, it remains a challenge to identify the complementary observables \cite{BMV3, BMV4, BMV5} that need to be measured on the subsystems to construct an entanglement witness. We also remark that the transfer of entanglement effects between the primordial entangled pair and other interacting species that scatter off the pair or interact gravitationally with it could be witnessed through the correlation functions of the scattered particles. We emphasize the importance of this effect as the creation of entanglement between particles whose interaction is mediated solely by gravity is a direct witness of the quantum nature of the gravitational field \cite{BMV1,BMV2}. 

Generalizing the above arguments to many body systems, one could consider a massive condensate of primordial particles, similar to axion stars. A class of witnesses predicated on the non-Gaussianity \cite{AVI1} of the state can be used as a probe of the quantum nature of the self-gravitational effects. 

Furthermore, we can consider a stationary state of two particles as discussed earlier. We may assume that the total state of the two-particle system is pure, which is reasonable given that there is limited decoherence from the environment. As an entanglement witness, we may measure the \( \tr \rho_1^2 \), where $\rho_1$ is the reduced density matrix of one of the particles (labelled as 1). We can then conclude that the particles are entangled if the value of \( \tr \rho_1^2 \) deviates significantly from unity.

\textit{Conclusion and Discussions} --- 
We have demonstrated that primordial particles can persist in a quantum state after interacting gravitationally with the dominant constituents of the universe at various epochs. The decoherence time of a primordial particle formed during recombination scales as \(M^2\) yielding an upper bound of \(10^7\, \mathrm{kg}\) on the mass of primordial particles that we expect to survive in a coherent state till present time. This is significant from the standpoint of quantum gravity where massive superpositions are expected to get entangled to the gravitational field probing its quantum nature. Our work provides a natural foray into the foundations of quantum gravity and helps ascertain the necessary postulates expected in candidate theories of quantum gravity. 

There are many topics which deserve to be explored further. For example, the difference between the interference patterns \(I_{\rm{cl}}\) and \(I_{\rm{qg}}\) introduced in \eqref{eq:cl-superpose} and \eqref{eq:qg-superpose} respectively can be calculated explicitly and in greater detail. Furthermore a comprehensive study of the nature and observability of the absorption lines must be conducted. 
We hope to address these issues in future work.

\section*{Acknowledgements}
We acknowledge several useful discussions with Chiara Marletto, Nicetu Tibau Vidal and Simone Rijavec. GAN and YW were supported in part by the NSFC Excellent Young Scientist (EYS) Scheme (Hong Kong and Macau) Grant No. 12022516. VV acknowledges funding from the National Research Foundation (Singapore), the Ministry of Education (Singapore), the Oxford Martin School, the John Templeton Foundation, the EPSRC (UK) and Wolfson College, University of Oxford. 

\appendix
\section{Decoherence of Massive Particle in a Photon Bath}
We study the decoherence rate of the model from \eqref{eqn:free-lagrangian} and \eqref{eqn:hint-pos}. We can introduce creation and annihilation operations \(a_\vk^\dg\) and \(a_\vk\) for \( \chi \), and \(b_\vk^\dg\) and \(b_\vk\) for photons respectively, and define \(\nu(\vk)\) to be the Fourier transform of gravitational potential \( \phi(\vx) = M/|\vx| \) (see Appendix B),
\begin{equation} \label{ft}
    \nu(\vk) = \frac{1}{2 \pi}\int \vd\vx \,  \ee^{-\i \vk \cdot \vx}\, \phi(\vx) = \frac{M}{\pi k^2}.
\end{equation} 
Then, 
\begin{align}
    H_\text{int} &= \int  \vd \vp \vd\vk\,  \nu(\vk)\, \epsilon_\vk\, a_\vp^\dg a_{\vp + \vk}\, b_\vk^\dg b_\vk.
\end{align}
We define the Fourier transformed number density of massive particle as,
\begin{equation}
    N_\vk = \int \vd\vp a_\vp^\dg a_{\vp + \vk}.
\end{equation}
Then,
\begin{equation}
    H_\text{int} = \int \vd\vk\,  \nu(\vk)\, \epsilon_\vk\, N_\vk\, b_\vk^\dg b_\vk.
\end{equation}

\textit{Evolution of density matrix} --- We now study the evolution of the reduced density matrix \(\rho\) of the massive particle to calculate the time of decoherence under this interaction. The master equation has been studied in Ref. \cite{Halliwell2007} and takes the Lindblad form given as follows,
\begin{align} \label{eqn:lindblad}
\td \rho  \Delta t &= \i [ \tr_\E (U_1 \rho_{\E}) - {\tr}_\E ( B \rho_\E), \rho ] + \nonumber \\
&\qquad \tr_{\E} \left(  U_1 \rho_T U_1 - \half U_1^2 \rho_T - \half \rho_T U_1^2 \right),
\end{align}
where \( B \) is an arbitrary Hermitian operator, which would not be relevant in the slow motion approximation where time evolution of \( N_\vk \) is neglected in comparison to the rapidly evolving environment, and \( U_1 \) is the time evolution operator defined by,
\begin{equation}
    U_1 = - \int_{0}^{\Delta t} \d t H_\text{int}(t) \approx - \Delta t \int \vd\vk\,  \nu(\vk)\, \epsilon_\vk\, N_\vk\, b_\vk^\dg b_\vk.
\end{equation}
Here, \( \Delta t \) is the timescale over which we study the evolution of \(\rho\). This timescale is small compared to the evolution of the system but large compared to the evolution of the environment. The right hand side of this equation is also proportional to \(\Delta t\) as calculated in Ref. \cite{Halliwell2007}. We then obtain the following differential equation describing \( \rho \),
\begin{align} \label{eqn:master-equation}
\td \rho &= -\i \Big[ H_0 - \int \vd \vk d(\vk) N_\vk N^\dg_\vk , \rho \Big] +  \nonumber \\ 
&\qquad \int \vd \vk c(\vk) \Bigg(  N_\vk \rho  N_\vk^\dg - \half  N_\vk^\dg N_\vk \rho - \half \rho  N_\vk^\dg N_\vk  \Bigg).
\end{align}
Here, \( H_0 \) is the free Hamiltonian of the massive particle, \( d(\vk) \) is some function of \( \vk \) which depends on the operator \(B\) in \eqref{eqn:lindblad}, and \( c(\vk) \) is a function describing the photon environment given by, 
\begin{align}
    c(\vk) &= \frac{1}{2 \pi} |\nu(k)|^2 \, \epsilon_\vk^2 \, \la b^\dg_\vk b_\vk b^\dg_\vk b_\vk \ra_\E  \nonumber \\
    &=\frac{1}{2 \pi} |\nu(k)|^2\, \epsilon_\vk^2 \,  n_\vk (n_\vk +1). 
\end{align}
Here, we have used the thermal density of photons \(n_\vk\) in place of \(  \la b^\dg_\vk b_\vk \ra \). Notice that the integral containing \( d(\vk) \) in \eqref{eqn:master-equation} is a c-number due to the slow motion approximation and its commutator with \(\rho\) vanishes. We further restrict our focus to the one particle sector for the massive particle. Then the density matrix in \eqref{eqn:master-equation} reduces to a function \( \rho(\vk, \vk') := \la \vk | \rho | \vk' \ra \) of two momenta, \( \vk \) and \( \vk' \). In this representation,
\begin{align}
    \la \vk | [ H_0, \rho ] | \vk' \ra &= 
    \Bigg\la \vk \Bigg| \frac{k^2}{2M}\, \rho - \rho\, \frac{k'^2}{2M}  \Bigg| \vk' \Bigg\ra \nonumber \\
    &= \frac{k^2 - k'^2}{2M}\, \rho(\vk,\vk').
\end{align}
Further, as \( N_\vk N^\dg_\vk \) acting on the one particle state is a c-number, the commutator of \( \int \vd \vk d(\vk) N_\vk N^\dg_\vk  \) with \( \rho \) vanishes for any function \( d(\vk) \). Likewise,
\begin{align}
  \la \vk |  N_\vq \rho  N_\vq^\dg | \vk' \ra &=  \la \vk - \vq |  \rho | \vk' - \vq \ra = \rho(\vk - \vq,\vk' - \vq), \text{ \&} \nonumber \\
  \la \vk |  N_\vq^\dg N_\vq \rho   | \vk' \ra &= \la \vk |  \rho N_\vq^\dg N_\vq    | \vk' \ra =  \la \vk |  \rho | \vk' \ra = \rho(\vk ,\vk').
\end{align}
Also, as \( \nu(\vk) \) only depends on the magnitude \(k\), we shall write \( \nu(k) \) to denote it. With these simplifications, \eqref{eqn:master-equation} becomes,
\begin{widetext}
\begin{equation} \label{eqn:one-particle-master}
    \td \rho(\vk,\vk') = -i\rho(\vk,\vk')\, \frac{k^2 - k'^2}{2M} +  \frac{1}{2\pi} \int \vd \vq \, |\nu(q)|^2 \, \epsilon_\vq^2\, n_\vq (n_\vq +1) \Big[ \rho(\vk-\vq,\vk'-\vq) - \rho(\vk,\vk') \Big].
\end{equation}

\textit{Master equation for \( \tr \rho^2 \)} --- We now derive a master equation for \( \tr \rho^2 \). First, in the function form \( \rho(\vk,\vk') \), \(\rho^2 \) is given by,
\begin{equation} \label{eqn:rho-sq}
    \rho^2(\vk,\vk') = \int \vd \vs \rho(\vk,\vs)\rho(\vs,\vk').
\end{equation}
The evolution of \( \tr \rho^2 \) is given by,
\begin{align}
    \td{(\tr\rho^2)}(\vk,\vk') & = \bigintsss \vd \vk \vd\vs \bigg(  \rho(\vk,\vs) \td{\rho(\vs,\vk)} 
    + \td{\rho(\vk,\vs)}\rho(\vs,\vk)  \bigg) \nonumber \\
    &= \frac{1}{2\pi}  \bigintsss \vd \vq \, |\nu(q)|^2 \, \epsilon_\vq^2\, n_\vq (n_\vq +1) \Bigg[ - 2\tr\rho^2 + 2 \Re \int \vd\vk \vd\vs  \rho(\vk,\vs) \rho(\vs-\vq,\vk-\vq) \Bigg].
\end{align}
\end{widetext}
For the purpose of the integrals over \( \vs \) and \(\vk \), the direction of \( \vq \) is arbitrary and can be chosen to be along the \(z\)-axis. In other words, the integral only depends on the magnitude of \( \vq \). We can define,
\begin{equation} \label{eqn:lambda}
    \Lambda (q) := \tr \rho^2 - \Re  \int \vd\vk \vd\vs  \rho(\vk,\vs) \rho(\vs-q\, \hat{z},\vk-q\, \hat{z}).
\end{equation}
Then, 
\begin{equation}
\label{trrho2}
    \td{(\tr\rho^2)}(\vk,\vk') =  -\frac{1}{\pi} \int \vd \vq \, |\nu(q)|^2 \, \epsilon_\vq^2\, n_\vq (n_\vq +1)\, \Lambda (q).
\end{equation}
\textit{Bounds on \( \Lambda(q) \)} --- To calculate bounds on \( \Lambda(q) \) define,
\begin{equation}
    \alpha(q) := \int \vd\vk \vd\vs  \rho(\vk,\vs) \rho(\vs-q\, \hat{z},\vk-q\, \hat{z}) = \tr \rho\, \Tilde{\rho_q},
\end{equation}
where \( \Tilde{\rho_q}(\vs, \vk) := \rho(\vs-q\, \hat{z},\vk-q\, \hat{z})  \) is the ``displaced" density matrix.
It has been shown \cite{Inequality} that for any real symmetric \( n \times n \) matrix \(B\) and any arbitrary real \( n \times n \) matrix \( A \),
\begin{equation} \label{eqn:lambda-bounds}
    \sum_{i=1}^n \lambda'_i\, \mu_{n-i+1} \leq \tr(AB) \leq \sum_{i=1}^n \lambda_i\, \mu_i.
\end{equation}
Here, \( \lambda_i \), \(\lambda'_i\) and \(\mu_i\) denote the \(i\)-th eigenvalue of \(A\), the transpose of \(A\) and \(B\) respectively when arranged in ascending order. 

Now, let \(A = \rho \), and \( B = \Tilde{\rho_q} \). \(A\) and its transpose have the same set of eigenvalues, say \( \{ \lambda_i \} \), and let the set of eigenvalues of \(B\) be \( \{ \mu_i \} \). 
Further, \(A\) and \(B\) are both valid density matrices and their eigenvalues represent probability to be found in some pure quantum state. Thus, they must all be positive. Thus, the lower bound in \eqref{eqn:lambda-bounds} is 0. 
We note that \( \tr A^2 = \tr B^2 = \tr \rho^2 \), \( \tr A^2 = \sum_i \lambda_i^2 \) and \( \tr B^2 = \sum_i \mu_i^2 \). Then, the upper bound can be obtained using the Cauchy-Schwartz inequality as,
\begin{equation}
    \tr(AB) \leq \sum_{i=1}^n \lambda_i\, \mu_i \leq \sqrt{\Big( \sum_i \lambda_i^2  \Big) \Big( \sum_i \mu_{i}^2  \Big)} = \tr\rho^2.
\end{equation}
Thus, 
\begin{equation} \label{eqn:lambda-limits}
    \alpha(q) \in [0, \tr\rho^2].
\end{equation}
and,
\begin{equation} \label{eqn:alpha-limits}
    \Lambda(q) = \tr \rho^2 - \alpha(q) \in [0, \tr\rho^2].
\end{equation}
\textit{Evolution of \( \tr \rho^2 \)} --- All the terms in \eqref{trrho2} only depend on the scalar magnitude of \( \vq \). Thus, we can integrate out the solid angle to \( 4 \pi \), and substitute \( \nu \) from \eqref{ft} and the bounds on \( \Lambda(q) \) to get, 
\begin{equation}
    \td{(\tr\rho^2)}(\vk,\vk') = -\Gamma \tr \rho ^2,
\end{equation}
where the decay rate \( \Gamma \in [0,\Gamma_0] \), with,
\begin{align} \label{eqn:d-tr-rho2-final}
    \td{(\tr\rho^2)}(\vk,\vk') &= -\frac{4M^2}{\pi^2} \bigintsss_0^\infty \frac{\d q} {q^2} \epsilon_\vq^2 n_\vq ( n_\vq + 1 ).
\end{align}
For a thermal bath of photons, \(\epsilon_\vq = q\) is the photon energy and $n_\vq$ is to be the Planck number density,
\begin{equation} \label{eqn:planck}
    n_\vq = \frac{q^2}{\pi^2(\ee^{\beta q}-1)},
\end{equation}
in natural units with \( \beta = (k_B T)^{-1} \). However, this derivation holds for any species of thermal particles in the background with the right choice of \(n_\vq\) and \( \epsilon_\vq \), and combinations of them with \( \Gamma_0 \) being additive. 

\section{Fourier Transform of Gravitational Potential}
Here, we derive the Fourier transform of gravitational potential and prove \eqref{ft}. We have,
\begin{align}
\nu(\vk) &= \frac{1}{2 \pi} \int \vd \vx \, \ee^{-\i \vk \cdot \v{x}} \, \phi(\vx)  \nonumber \\
&= \frac{M}{2 \pi}\int_0^\infty \d x \, x\, \frac{\sin k x}{k x}.    
\end{align}
We see that this integral is oscillatory and thus does not converge. Hence, we modify the potential to include a Yukawa term \( \ee^{-\lambda x} \) for \( \lambda > 0 \) and set it to 0 after integration. That is,
\begin{equation}
    \phi(\vx) = \frac{M}{x}\, \ee^{-\lambda x}.
\end{equation}
Then, 
\begin{align}
\nu(\vk) &= \frac{M}{2 \pi} \int_0^\infty \d x \, x \int_{-1}^1 \d t \, \ee^{-\i kxt - \lambda x} \\\nonumber
&= \frac{M}{ \pi} \frac{1}{k^2 + \lambda^2} \stackrel{\lambda \to 0}{=}\frac{M}{\pi k^2},
\end{align}
as stated in \eqref{ft}.\\

\section{Decoherence due to Fermions}
We begin by computing a better bound for \(\Lambda(q)\) in place of \eqref{eqn:alpha-limits}. Observe that when \(q = 0\), \(~\rho_q = \rho\), \(\alpha(0)=\tr\rho^2\) and thus \(\Lambda(0)=0\). Also, as \(\Lambda(q\) is an even function its first derivative must vanish. Then, in the small \(q\) limit, it will be approximated by a Taylor expansion with leading term \( \propto q^2 \). To calculate this approximation, consider the density matrix expressed in terms of its eigenvectors \( | \psi_i \ra \) and corresponding eigenvalues \(\ld_i\) as,
\begin{equation}
    \rho = \sum_i \ld_i \poii.
\end{equation}
Then,
\begin{equation}
    \rho^2 = \sum_i \ld_i^2 \poii.
\end{equation}
Define \(S_q\) to be the shift operator in momentum \(\exp(\i q Z)\), which acts on momentum eigenstates as \( S_q | \vp \ra =| \vp + q \hat{z} \ra \). Here \(Z\) is the \(z\)-component of the position operator. Then,
\begin{equation}
    ~\rho_q = \sum_i \ld_i^2 S_q \poii S_q^{-1}.
\end{equation}
Now, \( \Ld(q) \) can be written in terms of \(\ld_i\), \( | \psi_i \ra \) and \( S_q \),
\begin{equation}
    \Ld(q) = \sum_{i,j,k} \ld_i \ld_j \la \psi_k \poii S_q \pojj S_q^{-1} \pk{k} - \sum_i \ld_i^2.
\end{equation}
We expand \(S_q\) to second order in \(q\) and use orthonormality of \( | \psi_i \ra \) to get,
\begin{align}
    \Ld(q) &= \half q^2 \sum_{i \neq j} \ld_i \ld_j |\pbi Z \pkj|^2 \nonumber \\
    &\leq \half q^2 \la x^2 \ra_\rho \leq q^2 \la x^2 \ra_\rho \tr\rho^2.
\end{align}
Here, \( \la x^2 \ra_\rho \) is the expectation value of \( x^2 \) under the density matrix \( \rho \). The last inequality is because, we are only interested values of \( \tr \rho^2 \) close to 1. This, approximation however is only valid in the low \(q\) limit, \( q \lesssim 1/\sqrt{\la x^2 \ra_\rho \tr\rho^2} \). Let \(D := \la x^2 \ra_\rho \tr\rho^2 \). Then, \eqref{eqn:general-gamma-0} for decoherence due to fermions of mass \( m \) becomes,
\begin{align}
    \Gamma_0 &= \frac{4M^2 m^2}{\pi^2} \Bigg[ \bigintssss_0^{\frac{1}{\sqrt{D}}} \d q D n_\vq (n_\vq + 1) + \nonumber\\ &\qquad \qquad \qquad \qquad \qquad \bigintssss_{\frac{1}{\sqrt{D}}}^\infty \frac{\d q}{q^2} n_\vq (n_\vq + 1) \Bigg].
\end{align}

\bibliographystyle{unsrt}
\bibliography{references}

\end{document}